\documentclass[onecolumn]{article}

\input{epsf.sty}
\usepackage{epsfig}
\usepackage{multicol}
\usepackage{array}
\usepackage{amsfonts,amsmath,amssymb}

\begin{document}
\begin{center}
{\Large{\bf Intensity and polarization of radiation reflected from  accretion disc}}\\
\medskip
{N. A. Silant'ev\thanks{E-mail: nsilant@bk.ru} , G. A. Alekseeva, V. V. Novikov}
\medskip
\end{center}
\begin{center}
 {Central Astronomical Observatory at Pulkovo of Russian Academy of Sciences,\\ 196140,
Saint-Petersburg, Pulkovskoe shosse 65, Russia\\}
\end{center}
\medskip
\begin{center}
{received .......2018,\qquad accepted.....}
\end{center}

\begin{abstract}
We consider the reflection of non-polarized radiation from the point-like sources above the accretion discs both the optically thick and optically thin. We investigate the dependence of the polarization  of reflected radiation on the aperture angle  of incident radiation.  The aperture angle is determined by the radius of accretion disc and the height of the source  above the disc. 
For optically thick accretion discs we  show that, if the aperture angle is  smaller 70 grad, then  the wave electric field oscillations of reflected radiation  are  parallel to the accretion disc plane.   For aperture angle greater than 70 grad the 
 electric field oscillations are parallel to the plane  "normal to accretion disc  - the line of sight".  The latter also holds  for  reflection from the optically thin accretion disc  independent of the aperture angle value.  
 
\end{abstract}

$^1$ E-mail: nsilant@bk.ru

\section{Introduction}

The black holes (BH) demonstrate the presence of accretion discs both optically thick and optically thin. About 300  supermassive black holes (SMBH) exist in active galactic nuclei (AGN). There are $\sim$ 20 BH in X-ray stellar binary systems ( see Cherepashchuk 2006).

The observed  linear polarization from (AGN)  in Seyfert galaxies frequently demonstrate two types of polarization. In the first type the wave electric field ${\bf E}$-oscillations  are parallel to the plane of accretion disc. In the second type  the  ${\bf E}$-oscillations  are  parallel to  the plane (${\bf nN})$, where ${\bf N}$ is the normal to the disc plane and ${\bf n}$ is the line of sight. It appears the same effect is  in X-ray binary systems ( see, for example, Fabrika 2004 and references therein). The most often observed polarization is parallel to the disc plane. In Seyfert-1 AGNs  the polarization frequently corresponds to oscillations in the plane (${\bf nN}$), but there are polarizations with intermediate values of position angles. Note that the measurement of linear polarization from AGN is one of very effective technique to obtain the information about the structure of AGN. 
 The polarimetric results are given in many papers (see, for example, Impey et al.1991; Goodrich \& Miller 1994; Smith et al. 2002, 2004; Marin 2014).  

   The structure of accretion discs is considered both without the relativistic corrections and with them (see Shakura \& Sunyaev 1973; Novikov \& Thorne 1973; Riffert \& Herold 1995; Hubeny I. \&  Hubeny V. 1997 etc.). The standard nonrelativistic model (Shakura \& Sunyaev 1973; Urpin 1983; Pariev \& Colgate 2007) is most popular. It should be noted that the models of accretion discs  demonstrate  the large difference in temperature and  density as the function of the distance $r$ from the center of a disc.

  The presence of different types of jets is widely adopted in literature (see, for example, Blandford \& Konigle 1979; the review of Blandford
1990; Blandford 2003; Hawley et al. 2007; Krivosheyev et. al. 2009). The various models of jets give rise to the spectrum of radiation - from X-ray up to optical region. These radiations escape from different heights in the jet. The mechanisms of jets can be different. The main problem of construction of jets models is how jet arises in the center of accretion disc. Usually one takes into  account the  influence of magnetic field and the velocity of accretion discs rotation. Besides, it is necessary to use the observed  radiation from the jet (if they are observed).

 The goal of our paper is investigation of the radiation reflection  from  accretion disc both optically thick and optically thin.  Note that the height $H$ of considered radiation source  depends on wavelength.  We demonstrate that the intensity and  polarization of reflected  radiation  depend on aperture angle $\Theta$  of the incident radiation. We also present the emerging radiation going from optically thick layers of the accretion disc ( the Milne problem). 

\section{Statement of problem}

 Chandrasekhar (1960) derived the radiative transfer equation  for the intensity $I(\tau,\mu)$ and the Stokes parameter $Q(\tau,\mu)$ for an atmosphere consisting of free electrons. This equation, which we use  in our calculations,  has the matrix phase function  $\hat P(\mu,\mu')$. Here $\mu={\bf nN}=\cos\theta,\, \mu'={\bf n'N}=\cos\theta'$,  where the angles $\theta$ and $\theta'$ characterize the line of  sight ${\bf n}$ and the direction of incident radiation ${\bf n}'$, respectively (see Fig.1).
 The factorization of the matrix phase function $\hat P(\mu,\mu')$, i.e. the presentation  of this matrix as a product of two matrices $\hat P(\mu,\mu')=\hat A(\mu)\hat A^T(\mu')$, plays very important role in radiative transfer theory. The subscript $T$ stands for the matrix transpose.  We use the factorization presented in many papers (see, for example, Ivanov 1995; Silant'ev et al. 2017).   Note that we restrict ourselves by conservative axially symmetric atmosphere.

\begin{figure*}[h!]
\fbox{\includegraphics[ width=16cm, height=10cm]{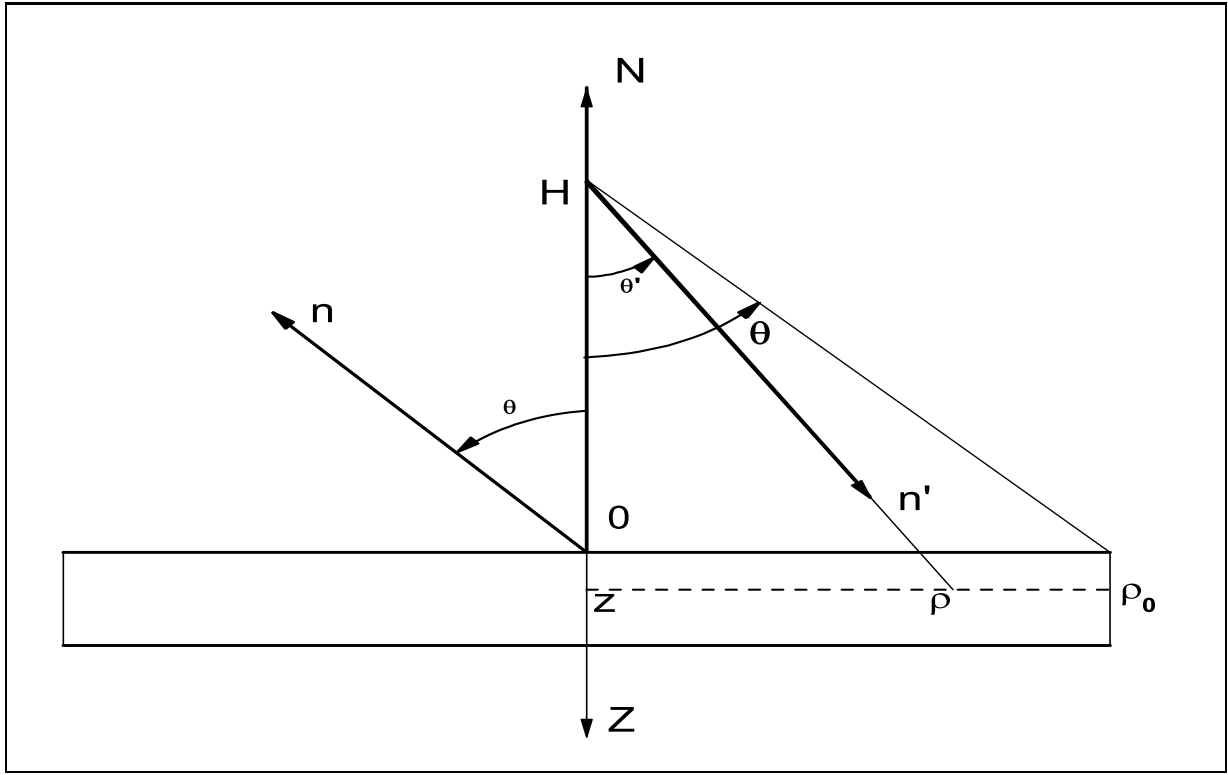}}
\caption{The basic notions}
\label{a}
\end{figure*}

  The radiative transfer equation for the vector (column) with the components $(I,Q)$ can be written in the form:
 \[
({\bf n}\nabla){\bf I}({\bf r},\mu)= -\alpha{\bf I}({\bf r},\mu)+ \alpha\hat{A}(\mu){\bf  K}({\bf r}) +
\]
\begin{equation}
 \alpha\hat A(\mu)\hat s({\bf r})\left (\begin {array}{c} 1 \\ 0 \end{array} \right ),
\label{1}
\end{equation}
\noindent where $\alpha$ is the extinction factor. We use the cylindrical frame of reference ${\bf r} =(z,\rho, \varphi)$.  The source term describes the single  scattered  incident radiation from the point-like source located above the accretion disc. Recall, that we consider single  or more scattered light, i.e.  the diffused radiation (see Chandrasekhar 1960).

 For  point-like source of non-polarized  light with the aperture of incident photon $\theta'$ ($\mu'=\cos\theta' =(z+H)/\sqrt{(z+H)^2+\rho^2} $) we have:

\begin{equation}
\hat s({\bf r})=L_0(\mu')\frac{\exp{( -\tau/\mu')}}{\rho^2+(z+H)^2}\hat A^T(\mu').
\label{2}
\end{equation}  $L_0(\mu')$ is the luminosity of the source. In general case the luminosity can  be anisotropic. The value  $H$ is the height of the source above the surface of accretion disc. {\bf The axis $Z$ is  directed inside the accretion disc.} The value $z=0$ corresponds to the surface  of accretion disc.

 In our case the source has the spot-like form. Far from this "spot" the intensity $I$ and parameter $Q$ tend to zero. A telescope is located far from the accretion disc and observes the integral radiation.

 The accretion disc, as a whole, looks like point source of radiation.  The integration over all surface ($2\pi\rho d\rho $)  gives rise to usual radiative transfer equation for these integrated values (we use the same notions for them).

 Introducing the optical depth $d\tau=\alpha dz$, we obtain the equation:
 \begin{equation}
\mu\frac{d{\bf I}(\tau,\mu)}{d\tau}= - {\bf I}(\tau,\mu)+ \hat{A}(\mu){\bf  K}(\tau) + \hat A(\mu)\hat s(\tau)\left (\begin {array}{c} 1 \\ 0 \end{array} \right ).
\label{3}
\end{equation}
\noindent Eq.(3) is appears as  the usual radiative transfer equation in plane parallel atmosphere (Sobolev 1963).

The vector ${\bf K}(\tau)$ has the form:
\begin{equation}
{\bf K}(\tau)=\frac{1}{2}\int_{-1}^1\,d\mu\hat A^T(\mu){\bf I}(\tau,\mu).
\label{4}
\end{equation}

  The factorization  matrix $\hat A(\mu)$ ( see Ivanov 1995; Silant'ev et al. 2017) is equal to:
\begin{equation}
\hat A(\mu)= \left (\begin{array}{rr}1 ,\,\, \,\sqrt{C}(1-3\mu^2) \\ 0 ,\,\,\,3\sqrt{C} (1-\mu^2) \end{array}\right).
\label{5}
\end{equation}
\noindent Here $C=1/8=0.125$. 
 
 Using the relation $\rho d\rho/[(\rho^2+(z+H)^2)]$=\,-\,$d\mu'/\mu'$, we obtain the following formula for $\hat s(\tau)$:
\begin{equation}
\hat s(\tau)=\int_{\cos\Theta}^1\frac{d\mu'}{\mu'}L_0(\mu')\hat A^T(\mu')\exp{(-\tau/\mu')}.
\label{6}
\end{equation}
\noindent The value $\Theta $ is the total aperture of incident radiation,  depending on the radius $\rho_0$ of accretion disc and the height $H$ of the point-like source: $\cos\Theta=H/\sqrt{H^2+\rho_0^2}$. If the radius of the accretion disc tends to infinity, then $\Theta \to  90^{\circ}$. This case is considered in the papers Grinin \& Domke (1971) and Silant'ev \& Gnedin (2008). In this case the source $\hat s(\tau)$ is independent of $H$.  For accretion discs in X-ray stellar binary systems the radius $\rho_0$ is  the distance between the components.  Note that reverberation technique, used by Fabian, Zoghbi, Ross et al.(2009a), (2009b); Zoghbi \& Fabian (2011), gave the estimation of the height $H\simeq 10 r_g$ for the sources of X-ray radiation.  Here $r_g$ is the gravitational  radius of black hole.  Such small $H$ gives the $\Theta\simeq 90^{\circ}$. Recall, that in this case  the reflected radiation is independent of the height $H$ and the  polarization corresponds to ${\bf E}$-oscillations in the plane $({\bf nN})$.

 In principle, the total aperture can be  determined as the angle  inside of which  all the radiation falls on the disc.  

\section{ The solution by resolvent technique}

The radiative transfer equation (3) can be written in more compact form:
 \begin{equation}
\mu\frac{d{\bf I}(\tau,\mu)}{d\tau}= - {\bf I}(\tau,\mu)+ \hat{A}(\mu){\bf  S}(\tau),
\label{7}
\end{equation}
\begin{equation}
{\bf S}(\tau)={\bf K}(\tau)+\hat s(\tau)\left (\begin {array}{c} 1 \\ 0 \end{array} \right ).
\label{8}
\end{equation}

Using the formal solution of Eq.(7), we can derive the integral equation for ${\bf S}(\tau)$:

\begin{equation}
{\bf S}(\tau)=\hat s(\tau)\left (\begin {array}{c} 1 \\ 0 \end{array} \right )+\int_0^{\infty}d\tau'\hat L(|\tau-\tau'|){\bf S}(\tau'),
\label{9}
\end{equation}
\noindent where the matrix kernel $\hat L(|\tau-\tau'|)$ has the form:
\begin{equation}
\hat L(|\tau-\tau'|)=\int_0^1\frac{d\mu}{\mu}\hat \Psi(\mu)\exp{(-|\tau-\tau'|/\mu)},
\label{10}
\end{equation}
\begin{equation}
\hat \Psi(\mu)=\frac{1}{2}\hat A^T(\mu)\hat A(\mu)\equiv\hat \Psi^T(\mu).
\label{11}
\end{equation}
The general theory for calculation of  the vector ${\bf S(\tau)}$ is  given in Silant'ev et al.(2015).
    According to this theory, Eq.(9) has solution through the resolvent matrix  $\hat{R}(\tau,\tau')$ (see Smirnov (1964), Sobolev (1969)):
\begin{equation}
{\bf S}(\tau)=\hat s(\tau)\left (\begin {array}{c} 1 \\ 0 \end{array} \right )+\int_0^{\infty}d\tau'\hat R(\tau,\tau')
\hat s(\tau')\left (\begin {array}{c} 1 \\ 0 \end{array} \right ),
\label{12}
\end{equation}
\noindent  where the resolvent matrix obeys the integral equation:
\begin{equation}
\hat{R}(\tau,\tau')=\hat L(|\tau-\tau'|)+\int_0^{\infty}d\tau'\hat L(|\tau-\tau''|)\hat R(\tau'',\tau').
\label{13}
\end{equation}

 $\hat R(\tau,\tau')$ can be calculated  if we know the matrices $\hat{R}(\tau,0)$ and $\hat{R}(0,\tau)$.
The kernel $\hat{L}(|\tau-\tau'|)$ of equation for $\hat{R}(\tau,\tau')$ is symmetric: $\hat{L}=\hat{L}^T$ . This gives rise to the relation  $\hat{R}(\tau,\tau')=\hat{R}^T(\tau',\tau)$.
 Note (see Silant'ev et al.(2015)), that the double Laplace transform of  $\hat R(\tau,\tau')$ has the form:
\[
\tilde{\tilde{\hat{R}}}\left(\frac{1}{\mu},\frac{1}{\mu'}\right)=\frac{\mu\mu'}{\mu+\mu'}\left [\,\tilde{\hat{R}}\left(\frac{1}{\mu},0\right)+\right.
\]
\begin{equation}
\left. \tilde{\hat{R}}\left(0,\frac{1}{\mu'}\right)+\tilde{\hat{R}}\left(\frac{1}{\mu},0\right)
\tilde{\hat{R}}\left(0,\frac{1}{\mu'}\right)\right].
\label{14}
\end{equation}
The Laplace transform of $R(\tau,0)$ with the  parameter $1/\mu$ plus the unit matrix $\hat E$  is known  as $\hat H(\mu)$ - matrix:
\begin{equation}
\hat H(\mu)=\hat E+\tilde{\hat{R}}\left(\frac{1}{\mu},0\right).
\label{15}
\end{equation}
From Eqs.(14) and (15) we obtain the following  nonlinear equation:
\begin{equation}
\hat H(\mu)=\hat E+\mu\int_0^1d\mu'\frac{\hat H(\mu)\hat H^T(\mu')\hat \Psi(\mu')}{\mu+\mu'}.
\label{16}
\end{equation}

\section{The emerging radiation from optically thick accretion discs}

 According to Eq.(7) the emerging radiation ${\bf I}(0,\mu)$  has the form:

\[
{\bf I}(0,\mu)=\hat A(\mu)\int_0^{\infty}\frac{d\tau}{\mu}{\bf S}(\tau)\exp{(-\tau/\mu)}=
\]
\begin{equation}
\hat A(\mu)\int_{\cos\Theta}^1d\mu'\frac{L_0(\mu')\hat H(\mu)\hat H^T(\mu')\hat A^T(\mu')}{\mu+\mu'}\left (\begin {array}{c} 1 \\ 0 \end{array} \right ).
\label{17}
\end{equation}
\noindent Recall, that $\mu=\cos\theta $, where $\theta$ is the angle between the line of sight ${\bf n}$ and the normal ${\bf  N}$ to  the accretion disc, i.e. this is the inclination angle. We emphasize that ${\bf I}(0,\mu)$ depends on the aperture $\Theta$ of incident radiation.

For  numerical calculations it is better introduce the new matrix $\hat D(\mu)=\hat A(\mu)\hat H(\mu)$, which obeys the equation:

\[
\hat{D}(\mu)\equiv \left (\begin{array}{rr}a(\mu) ,\,\, b(\mu) \\ c(\mu),\,\,\,d(\mu) \end{array}\right)=
\]
\begin{equation}
\hat A(\mu)+
\frac{\mu}{2}\int_0^1d\mu'\frac{\hat D(\mu)\hat D^T(\mu')\hat A(\mu')}{\mu+\mu'}.
\label{18}
\end{equation}
\noindent The kernel of this equation does not depend on $\mu'^4$, i.e. Eq.(18) is simpler than Eq.(16). The technique of numerical calculation of $\hat D(\mu)$ is given in Silant'ev et al.(2017). Expression (17) in new matrix takes the form:
\begin{equation}
{\bf I}(0,\mu)=\int_{\cos\Theta}^1d\mu'\frac{L_0(\mu')\hat D(\mu)\hat D^T(\mu')}{\mu+\mu'}\left (\begin {array}{c} 1 \\ 0 \end{array}. \right )
\label{19}
\end{equation}
\noindent  It is of  interest, that directly from Eq.(18)   one can obtain that zero's moments $a_0=2$ and $b_0=0$ (see Silant'ev et al. 2017).
 The numerical calculation confirms this.
 The functions $a(\mu), b(\mu), c(\mu)$ and $d(\mu)$  are given in Table 1.

\begin{table}[h]
\caption {\small  Functions $a(\mu), b(\mu), c(\mu), d(\mu)$.}
\scriptsize
\begin{tabular}{|p{0.5cm} | p{0.9cm} p{1.2cm} p{0.9cm} p{0.9cm}| }
\hline
\noalign{\smallskip}
$\mu$ & $a(\mu)$ & $ b(\mu)$ & $c(\mu)$ & $ d(\mu)$   \\
\hline
\noalign{\smallskip}
0       & 1           & 0.3536   &  0            & 1.0607           \\
0.01 & 1.0370  & 0.3706   & 0.0075    & 1.0912      \\
0.02 & 1.0670  & 0.3823   & 0.0126    & 1.1128    \\
0.03 & 1.0946  & 0.3919   & 0.0170    & 1.131     \\
0.04 & 1.1208  & 0.4001   & 1.0208    & 1.1472      \\
0.05 & 1.1460  & 0.4071   & 0.0244    & 1.1616        \\
0.06 & 1.1704 & 0.4133    & 0.0276    & 1.1746        \\
0.07 & 1.1941 & 0.4186    & 0.0306    & 1.1864        \\
0.08 & 1.2174 & 0.4231    & 0.0334    & 1.1971        \\
0.09 & 1.2402 & 0.4270    & 0.0361    & 1.2070       \\
0.10 & 1.2626 & 0.4303    & 0.0386    & 1.2159      \\
0.15 & 1.3703 & 0.4378    & 0.0492    & 1.2493     \\
0.20 & 1.4728 & 0.4325    & 0.0574    & 1.2667       \\
0.25 & 1.5716 & 0.4155    & 0.0637    & 1.2702        \\
0.30 & 1.6675 & 0.3874    & 0.0683    & 1.2611       \\
0.35 & 1.7611 & 0.3486    & 0.0714    & 1.2400      \\
0.40 & 1.8526 & 0.2995    & 0.0731    & 1.2075     \\
0.45 & 1.9422 & 0.2401    & 0.0735    & 1.1639     \\
0.50 & 2.0302 & 0.1705    & 0.0726    & 1.1096      \\
0.55 & 2.1166 & 0.0909    & 0.0705    & 1.0446     \\
0.60 & 2.2015 & 0.0014    & 0.0672    & 0.9692     \\
0.65 & 2.2851 & -0.0981   & 0.0627    & 0.8834    \\
0.70 & 2.3673 & -0.2074   & 0.0571    & 0.7874       \\
0.75 & 2.4482 & -0.3266   & 0.0503    & 0.6812       \\
0.80 & 2.5279 & -0.4555   & 0.0425    & 0.5650      \\
0.85 & 2.6063 & -0.5943   & 0.0335    & 0.4387      \\
0.90 & 2.6835 & -0.7428   & 0.0234    & 0.3024       \\
0.91 & 2.6988 & -0.7737   & 0.0213    & 0.2739      \\
0.92 & 2.7140 & -0.8050   & 0.0191    & 0.2451       \\
0.93 & 2.7293 & -0.8366   & 0.0168    & 0.2158      \\
0.94 & 2.7444 & -0.8686   & 0.0146    & 0.1862      \\
0.95 & 2.7595 & -0.9011   & 0.0122    & 0.1561     \\
0.96 & 2.7746 & -0.9339   & 0.0099    & 0.1257     \\
0.97 & 2.7896 & -0.9671   & 0.0075    & 0.0949      \\
0.98 & 2.8046 & -1.0007   & 0.0050    & 0.0636     \\
0.99 & 2.8195 & -1.0347   & 0.0025    & 0.0320     \\
1      & 2.8344 & -1.0691   & 0.0000    & 0.0000        \\
\hline 
\end{tabular}
\end{table} 

 A telescope  observes the direct radiation flux  from the source $F_{direct}=L_0(\mu)/R^2$, where $R$ is the distance to the telescope. The flux of scattered radiation is equal to  ${\bf F}_{diff}(\mu)=\mu{\bf I}(0,\mu)/R^2$.

 Note  that there exists the relation:
\begin{equation}
\left (\begin {array}{c} 1 \\ 0 \end{array} \right )=\hat A(\mu) \left (\begin {array}{c} 1 \\ 0 \end{array} \right ).
\label{20}
\end{equation}
\noindent Substitution relation (20) in Eq.(19) gives rise to the formula for ${\bf F}_{diff}(\mu)$:
\[
{\bf F}_{diff}(\mu)=
\]
\begin{equation}
\frac{\mu}{R^2}\int_{\cos\Theta}^1d\mu'\frac{L_0(\mu')\hat D(\mu)\hat D^T(\mu')\hat A(\mu')}{\mu+\mu'}
 \left (\begin {array}{c} 1 \\ 0 \end{array} \right ).
\label{21}
\end{equation}
\noindent According to Eq.(18) for isotropic source ($L_0(\mu)=L_0$) and the aperture $\Theta=90^{\circ}$  formula (21) takes the simple form:

\begin{equation}
{\bf F}_{diff}(\mu)=\frac{2 L_0}{R^2}\left(\hat D(\mu)-\hat A(\mu)\right)\left (\begin {array}{c} 1 \\ 0 \end{array} \right ).
\label{22}
\end{equation}
\noindent Below we restrict ourselves by the isotropic source and different values of aperture $\Theta$. We present the total fluxes
$F_I(\mu)=F_{I(diff)}(\mu)+F_{I(direct)}$ and $F_Q(\mu)=F_{Q(diff)}(\mu)$ ($F_Q(0)=0$) without the common factor $L_0/R^2$.  For this reason $F_I(0)=1$ and $F_Q(0)=0$, because at $\mu=0$ the flux ${\bf F}$ consists of the non-polarized direct radiation from the source.

According to Eq.(22), the total $F_I(\mu)$ and $F_Q(\mu)$ can be given in the form:
\[
F_I(\mu)=\frac{L_0}{R^2}(2a(\mu)-1),
\]
\begin{equation}
F_Q(\mu)=\frac{2L_0}{R^2}\,c(\mu), \,\,\, F_I(0)=\frac{L_0}{R^2}.
\label{23}
\end{equation}

 In Table 2 we give the
angular dependence $J(\mu)=F_I(\mu)$ and the polarization degree $p(\mu)=F_Q(\mu)/F_I(\mu)$ in \% for aperture $\Theta=90^{\circ}$.  Note that $p(\mu)$ for $\Theta=90^{\circ}$ is positive. This corresponds to the wave  electric field oscillations parallel to the plane $(\bf nN)$ (see Table 2). 
For comparison we give the angular dependence $J_M(\mu)$ and polarization degree $p_M(\mu)$ for the Milne problem:
\begin{equation}
J_M(\mu)=\frac{a(\mu)+s\,b(\mu)}{a(0)+s\,b(0)},\,\,\,\,p_M(\mu)=\frac{c(\mu)+s\,d(\mu)}{a(\mu)+s\,b(\mu)},
\label{24}
\end{equation}
\noindent  where $s=-0.10628$ is the solution of homogeneous equation for ${\bf K}(0)$ (see Silant'ev et al. 2017)

\begin{table}[h]
\caption { \small  Functions $J_M(\mu)$, $ - p_M(\mu)$\%  for $\Theta=90^{\circ}$.}
\scriptsize
\begin{tabular}{|p{0.5cm} | p{0.9cm} p{1.2cm} p{0.9cm} p{0.9cm}| }
\hline
\noalign{\smallskip}
$\mu$ & $J_M(\mu)$ & $ - p_M(\mu)$ & $J(\mu)$ & $ p(\mu)$    \\
\hline
\noalign{\smallskip}
0         &  1         & 11.713  & 1         & 0           \\
0.01   & 1.037   & 10.875  & 1.074  & 1.388     \\
0.02   & 1.066   & 10.294  & 1.134  & 2.223   \\
0.03   & 1.094    & 9.804  & 1.189  & 2.854     \\
0.04   & 1.120    & 9.373  & 1.242  & 3.358      \\
0.05   & 1.146    & 8.985  & 1.292  & 3.772       \\
0.06   & 1.170    & 8.631  & 1.341  & 4.118       \\
0.07   & 1.194    & 8.304  & 1.388  & 4.411      \\
0.08   & 1.218    & 8.000  & 1.435  & 4.661      \\
0.09   & 1.241    & 7.716  & 1.480  & 4.876      \\
0.10   & 1.264    & 7.449  & 1.525  & 5.060    \\
0.15   & 1.375    & 6.312  & 1.741  & 5.656   \\
0.20   & 1.482    & 5.410  & 1.946  & 5.904      \\
0.25   & 1.587    & 4.667  & 2.143  & 5.945      \\
0.30   & 1.690    & 4.040  & 2.335  & 5.851     \\
0.35   & 1.791   & 3.502  & 2.522   & 5.663    \\
0.40   & 1.892    & 3.032  & 2.705  & 5.406   \\
0.45   & 1.991    & 2.619  & 2.884  & 5.096   \\
0.50   & 2.091    & 2.252  & 3.060  & 4.745    \\
0.55   & 2.189    & 1.923  & 3.233  & 4.361   \\
0.60   & 2.287    & 1.626  & 3.403  & 3.949   \\
0.65  & 2.385    & 1.358  & 3.570   & 3.513   \\
0.70  & 2.483    & 1.113  & 3.735   & 3.058     \\
0.75  & 2.580    & 0.888  & 3.896   & 2.584     \\
0.80  & 2.677    & 0.682  & 4.056   & 2.094     \\
0.85  & 2.774    & 0.492  & 4.213   & 1.590    \\
0.90  & 2.870    & 0.316  & 4.367   & 1.072     \\
0.91  & 2.890    & 0.282  & 4.398   & 0.967    \\
0.92  & 2.909    & 0.249  & 4.428   & 0.862     \\
0.93  & 2.928    & 0.216  & 4.458   & 0.756    \\
0.94  & 2.947    & 0.184  & 4.489   & 0.649    \\
0.95  & 2.967    & 0.152  & 4.519   & 0.542   \\
0.96  & 2.986    & 0.121  & 4.549   & 0.435   \\
0.97  & 3.005    & 0.090  & 4.579   & 0.327    \\
0.98  & 3.024    & 0.060 & 4.609    & 0.218   \\
0.99  & 3.044    & 0.030  & 4.639   & 0.109    \\
1       & 3.063    & 0        & 4.669     & 0           \\
\hline 
\end{tabular}
\end{table}

 Recall, that the Milne problem describes the diffusion of thermal radiation from the sources located far below  the surface of the optically thick accretion disc.
  In the Milne problem the polarization $p(\mu)$ is negative.  This corresponds to the wave electric field oscillations parallel to the accretion disc plane.  Note that the reflected radiation has the maximum polarization value $p=5.945$\% at $\theta=75.5^{\circ}$(see Table 2). This value is greater than corresponding absolute value in the Milne problem ($4.667$\%).  For $\theta=0$ and $\theta=90^{\circ}$ the polarization of reflected radiation desappears. The observed radiation intensity  is the sum of reflected intensity $F_Q(\mu)$ and the Milne problem intensity $F_M(\mu)$, escaping from the surface of the accretion disc. The observed polarization $p_{obs}(\mu)$ can be both positive and negative depending on the relative values of intensities $F_Q(\mu)$ and the Milne problem intensity $F_M(\mu)$. Tables 1 and 2 can be used for estimations of observed intensity and polarization for  different relative values of $F_Q(\mu)$ and $F_M(\mu)$ . So, these Tables allow us to estimate the intensities and polarization in different models. 

\begin{figure*}[h]
\fbox{\includegraphics [width=16cm]{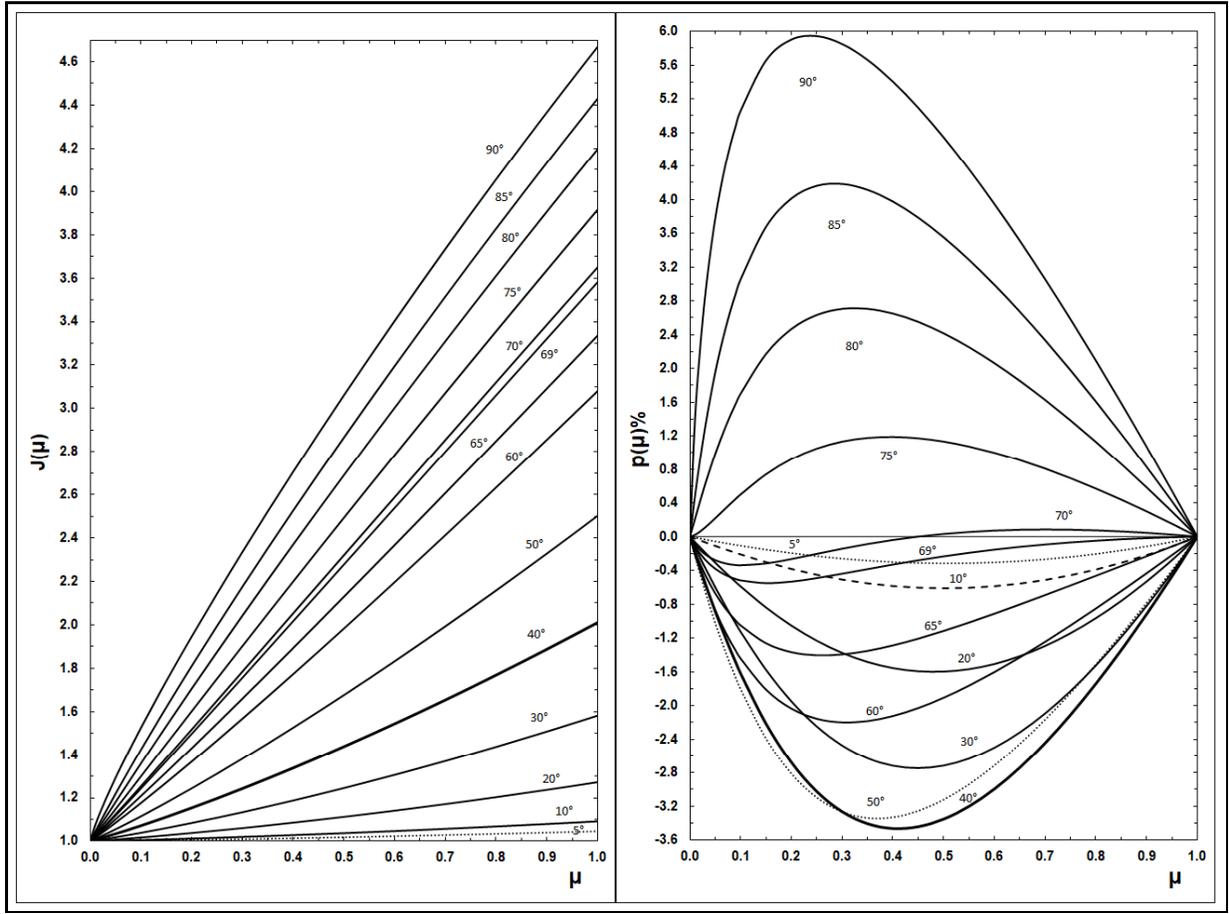}} 
\caption{\small The angular distribution $J(\mu)$ and polarization degree $p(\mu)$ in \% of observed radiation flux from optically thick accretion disc, consisting of free electrons. The numbers denote the apertures of point-like sources.}
\label{a}
\end{figure*}

Figure 2 presents the angular distribution of the emerging radiation $J(\mu)$ and the degree of polarization $p(\mu)$ in \% for a number of
values of aperture $\Theta$. We see that for $\Theta<70^{\circ}$ the ${\bf E}$ - oscillations are parallel to the accretion disc plane. The values $\Theta>70^{\circ}$ correspond to oscillations parallel to the plane $({\bf nN})$, which frequently occur in Seyfert-1 AGNs. Further we will explain  this behavior. 

\section{ The emerging radiation from optically thin accretion disc}

 The optically thin accretion disc is characterized by  small optical depth $\alpha z =\tau\ll 1$. In this situation we can neglect by multiple scattering of light. Instead of Eq.(17) for reflected radiation wave the expression:
\[
{\bf I}(0,\mu)=\hat A(\mu)\int_0^{\tau}\frac{d\tau}{\mu}{\bf s}(\tau)\exp{(-\tau/\mu)}=
\]
\[
\hat A(\mu)\int_0^{\tau}\frac{d\tau}{\mu}\exp{(-\tau/\mu)}\int_{\cos\Theta}^1\frac{d\mu'}{\mu'}L_0(\mu')\hat A^T(\mu')\times
\]
\begin{equation}
\exp{(-\tau/\mu')}\left (\begin {array}{c} 1 \\ 0 \end{array} \right ).
\label{25}
\end{equation}
\noindent Here we used the expression (6) for $ {\bf s}(\tau)$.  After integration over $d\tau$ we obtain the relation:
\[
{\bf I}(0,\mu)=\hat A(\mu)\int_{\cos\Theta}^1d\mu'L_0(\mu')\hat A^T(\mu')\times
\]
\begin{equation}
\frac{1-\exp{(-\tau(1/\mu'+1/\mu))}}{\mu+\mu'}\left (\begin {array}{c} 1 \\ 0 \end{array} \right ).
\label{26}
\end{equation}
\noindent The expression (26) takes  into account the single scattered radiation. The flux of radiation in a telescope is equal to:
\begin{equation}
{\bf F}(0,\mu)=\frac{1}{R^2}\left[\mu {\bf I}(0,\mu)+L_0(\mu)\left (\begin {array}{c} 1 \\ 0 \end{array} \right)\right],
\label{27}
\end{equation}
\noindent where $R$ is a distance to telescope.

 In Figs. 3 and 4 we give the angular distribution $J(\mu)$ and polarization degree $p(\mu)$ for accretion optical depths $\tau=0.1$ and $  0,3$ . Recall, that the scattered radiation intensity is proportional to optical depth $\tau$ and the value of aperture $\Theta$. In this
situation the intensity of direct radiation from the source is greater than the intensity of reflected radiation. The degree of polarization $p(\mu)$ in \% is more distinct. Note that all the forms of all Figures are similar. The difference exists with the  scales. It appears this is the consequence that they describe the single scattering in accretion discs.

\begin{figure*}[h]
\fbox{\includegraphics [width=16cm]{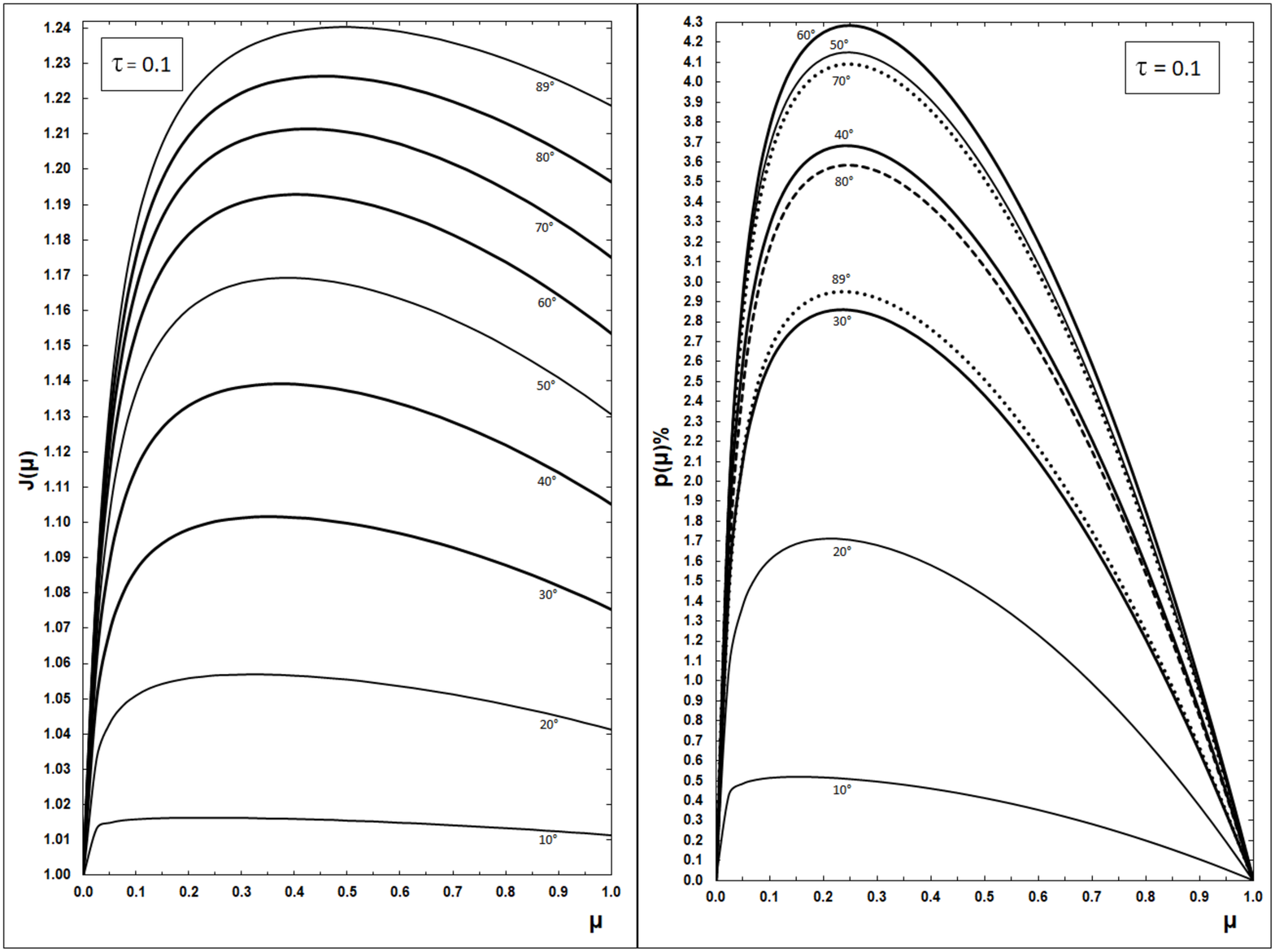}} 
\caption{\small The angular distribution $J(\mu)$ and polarization degree $p(\mu)$ in \% of observed radiation flux from optically thick accretion disc, consisting of free electrons. The numbers denote the apertures of point-like sources.}
\label{a}
\end{figure*}

\begin{figure*}[h]
\fbox{\includegraphics [width=16cm]{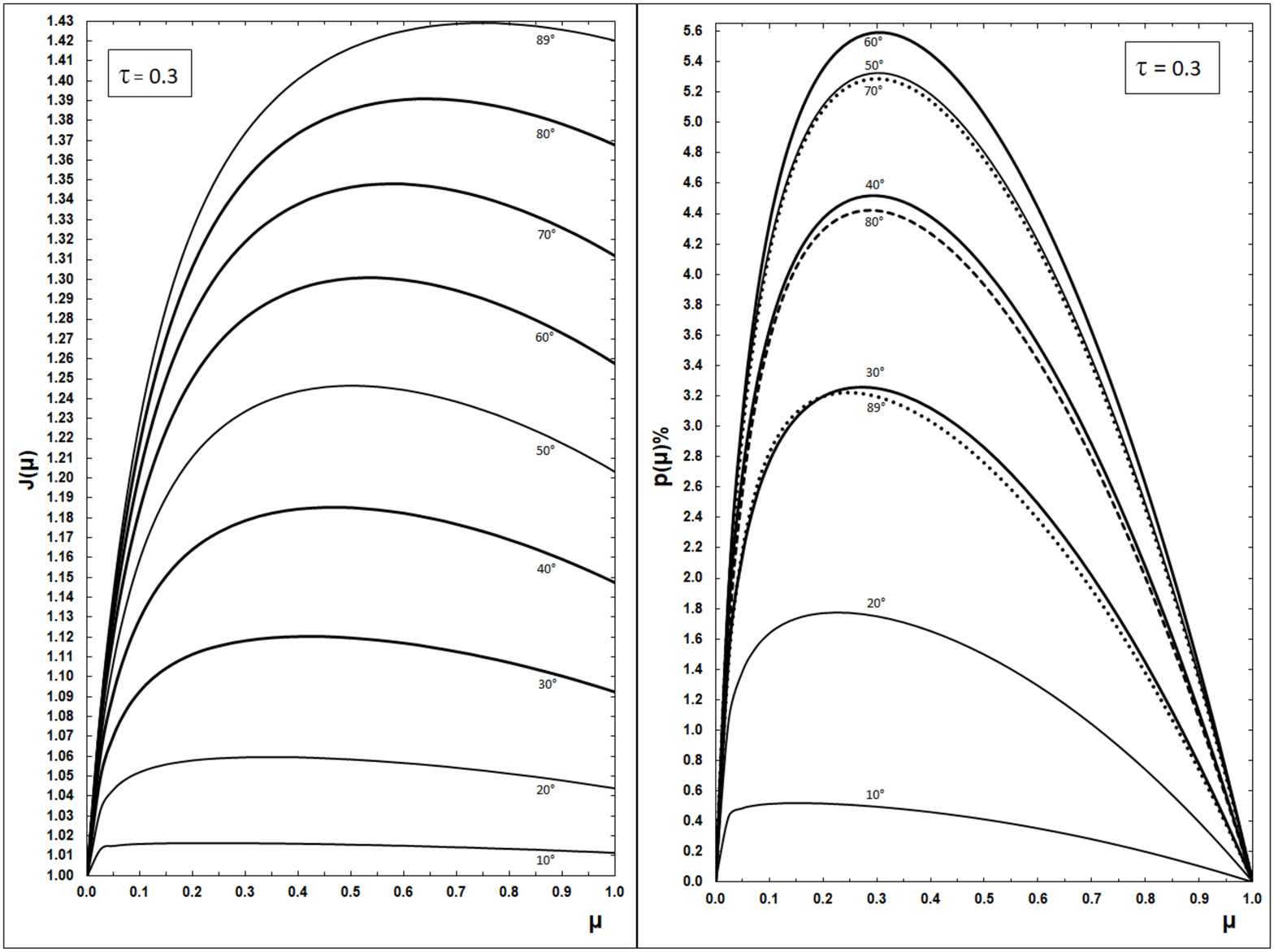}} 
\caption{\small The angular distribution $J(\mu)$ and polarization degree $p(\mu)$ in \% of observed radiation flux from optically thick accretion disc, consisting of free electrons. The numbers denote the apertures of point-like sources.}
\label{a}
\end{figure*}

\section{Discussion  of the calculation results }

 We  investigated the reflection of non-polarized radiation from point-like sources above the reflecting accretion disc. The optically thick and optically thin accretion discs were considered. We found  two types of polarization in reflected radiation. The first type corresponds to the wave electric field ${\bf E}$-oscillations parallel to the plane of accretion disc. The second type corresponds  to ${\bf E}$-oscillations parallel to the plane $({\bf nN})$, which is perpendicular to the accretion disc plane. Recall, that ${\bf n}$ is the line of sight direction, and the unit vector ${\bf N}$ is perpendicular to the accretion disc.

  Why arise  these types of polarization?  Note that the first type corresponds to Chandrasekhar's solution of the Milne problem. This type is named as the usual (positive) polarization. This polarization holds in many cases. The second type of polarization occurs rarely and was named as "negative" polarization.
Of course, there are promediate directions of the wave electric field oscillations between the positive and negative polarizations. They are characterized by concrete positional  angles.

Recall, that  ${\bf E}$-oscillations of single scattered non-polarized radiation are perpendicular to the plane of scattering $({\bf nn}_0)$, where ${\bf n}_0$ is the direction of  incident  radiation. If  ${\bf n}\perp{\bf n}_0$, then the scattered radiation is $100$\% polarized.
The degree of polarization $p(\mu_0)\sim \sin^2\mu_0$, where $\mu_0$ is the cosine of the angle between vectors ${\bf n}$ and ${\bf n}_0$.

  The case of optically thin accretion disc is most simple.  Let us consider two characteristic regions of scattering. In the first region the plane of scattering coincides with the plane $({\bf nN})$. Here the scattered radiation has polarization $\sim \sin\theta$ and the ${\bf E}$-oscillations  are parallel to the plane of accretion disc. In the second region the plane of scattering is perpendicular to plane $({\bf nN})$. Here the angle of  scattering radiation is $ 90^{\circ}$, i.e. the scattered radiation is $100$\% polarized and the ${\bf E}$-oscillations hold in the plane $({\bf nN})$.  As a result, for all $\Theta$ the total scattered radiation has the polarization parallel to the plane $({\bf nN })$ ("negative" polarization). In Figs.3 and 4  we give the result of calculations according to Eqs.(26) and (27). Note that the mentioned effects are weaker for small inclination angles $\Theta$.

In the first case  the non-polarized radiation from the  point-like source illuminates the optically thick accretion disc. Here the situation is more complex. The calculations demonstrate that for $\Theta <70^{\circ}$ the reflected radiation has polarization with ${\bf E}$-oscillations 
parallel to the accretion plane, i.e. is similar to the case of the Milne problem. That is because for these $\Theta $ the incident radiation penetrates deeper inside the optically thick accretion disc than  for radiation with the angles $\Theta >70^{\circ}$. With the increase of  $\Theta $  the most radiation does not penetrates in deep layers of accretion disc. The scattering of radiation became similar to single scattering case. As a result, the total emerging radiation has the polarization similar to the second type, considered above.

Note that the total radiation with the wavelength $\lambda$ can also  have other sources, not only  from a jet ( see, for example, Krivosheyev et al. 2009).

\section{Conclusion}
  In this paper we study the reflection of non-polarized radiation from the point-like sources located above the reflecting accretion disc. The optically thick and optically thin accretion discs are considered. We also take into account the intensity of radiation going to a telescope directly from  a source. For apertures $\Theta< 70^{\circ}$ the  polarization of total radiation,  reflected from optically thick disc, corresponds to wave electric field oscillations parallel to the plane of accretion disc. For apertures $\Theta>70^{\circ}$ the polarization corresponds to oscillations parallel to the plane (${\bf nN}$), where ${\bf  N}$ is the  normal to the disc and ${\bf n}$ is the line of sight. Such  polarization is  frequently observed in active galactic nuclei of Seyfert-1 galaxies. For optically thin accretion discs the 
reflected  radiation  has the polarization corresponding to the ${\bf E}$-oscillations parallel to the plane  $({\bf nN})$  independently of the aperture. We considered the accretion discs consisting of free electrons..

{\bf Acknowledgements.} 
This research was supported by the Program of Presidium of Russian Academy of Sciences N 28.

 We are very grateful to a referee for a number of useful remarks and advices.


\begin{thebibliography}{30}

\bibitem{1} Blandford, R. D., Konigle, A., Ap.J. {\bf 232}, 34 (1979)
\bibitem{2} Blandford, R. D.: Lecture Notes of Saas-Fee courses, Geneva Observatory, XII, 280pp. Springer Verlag, Heidelberg (1990).
\bibitem{3} Blandford, R. D.: AGN Jets. In Active galactic nuclei: from Central Engine to Host Galaxy. ASP Conference Series, Vol.290, 267 (eds. S.Collin, F. Combes, I. Shlosman) (2003)
\bibitem{4} Chandrasekhar, S.: Radiative transfer. Dover, New York (1960)
\bibitem{5} Cherepashchuk, A. M., Publ. Astron. Obs. Belgrade, N{\bf 80}, 3 (2006)
\bibitem{6}  Fabian, A. C., Zoghbi, A., Ross, R. R., et al., Nature, {\bf 459}, 540 (2009a)
\bibitem{7}  Fabian, A. C., Zoghbi, A., Ross, R. R., et al., arXiv: 0905.4383v1 (2009b)
\bibitem{8} Zoghbi, A.\& Fabian, A. C., MNRAS, {\bf 418}, 2542 (2011)
\bibitem{9} Fabrika, S., Astrophysics \& Space Physical Review, vol {\bf 12}, 1-100 (2004)
\bibitem{10} Goodrich, R. W., Miller, J. S., Ap.J.,{\bf 434}, 82 (1994)
\bibitem{11} Grinin, V. P., Domke, H., Astrophysics {\bf 7}, 211 (1971)
\bibitem{12} Hawley, J. F., Beckwith, K., Krolik, J. H., Astrophys. Space Sci., {\bf 311}, 117 (2007)
\bibitem{13} Hubeny I., Hubeny V., Ap. J. {\bf 484}, L37 (1997)
\bibitem{14} Impey, C. D. , Lawrence, C. R., Tapia, S., Ap.J., {\bf 375}, 46 (1991)
\bibitem{15} Ivanov V. V., A\&A, {\bf 303}, 609 (1995)
\bibitem{16} Krivosheyev, Yu. M., Bisnovatyi-Kogan, G. S., Cherepashchuk, A. M., Postnov, K. A., MNRAS, {\bf 394}, 1674  (2009)
\bibitem{17} Marin, F., MNRAS, {\bf 441}, 551 (2014)
\bibitem{18} Novikov, I. D., Thorne, K. S.: In Black Hole Astrophysics , eds. C. de Witt \& B. de Witt, Gordon \& Breach , New York (1973)
\bibitem{19} Pariev, V. I., Colgate S. A., Ap.J., {\bf 658}, 114 (2007)
\bibitem{20} Riffert, H., Harold, H., Ap. J., {\bf 450}, 508 (1995)
\bibitem{21} Shakura, N. I., Sunyaev R. A., A\&A, {\bf 24}, 337 (1973)
\bibitem{22} Silant'ev, N. A.\& Gnedin, Yu. N., A\&A, {\bf 481}, 217 (2008) 
\bibitem{23} Silant'ev, N. A., Alekseeva, G. A., Novikov, V. V., Astrophys. Space  Sci., {\bf  357}, 53 (2015)
\bibitem{24} Silant'ev, N. A., Alekseeva, G. A., Novikov, V. V., Astrophys. Space Sci., {\bf 362}, 151 (2017)
\bibitem{25} Smirnov, V. I.:  The course of higher mathematics, Vol. 4. Integral equations and Partial differential equations. Pergamon Press,  New York  (1964)
\bibitem{26} Smith, J. E., Young, S., Robinson, A., Alexander, Corbett, E. A., Giannuzzo, M.E.,  D. M., Axon, D. J., Hough, J. H.,  MNRAS, {\bf 335}, 773 (2002)
\bibitem{27} Smith, J. E., Robinson, A., Alexander, D. M., Young, S., Axon, D. J., Corbett, E. A., MNRAS, {\bf 350}, 140 (2004)
\bibitem{28}Sobolev, V. V.: A treatise on radiative transfer. Van Nostrand, N. Y. (1963)
\bibitem{29} Sobolev, V. V.: Course in theoretical astrophysics. NASA Technical Translation F-531, Washington (1969)
\bibitem{30} Urpin, V. A., Astroph. Sp. Sci., {\bf 90}, 79 (1983)
\end{thebibliography}
\end{document}